\documentclass[twocolumn,prb,showpacs,preprintnumbers,amsmath,amssymb]{revtex4}
\usepackage{graphicx}
\usepackage{dcolumn}
\usepackage{bm}
\usepackage{epsfig}

\begin{document}
\title{Breathers and Raman scattering in a two-leg ladder with staggered Dzialoshinskii-Moriya
  interaction}
\author{E.~Orignac}
\affiliation{Laboratoire de Physique de l'Ecole Normale Superieure de
  Lyon, CNRS UMR5672, 46 All\'ee d'Italie, F-69364 Lyon Cedex 07, France.}
\author{R.~Citro}
\affiliation{Dipartamento di Fisica ``E. R. Caianiello''and Unit\`a
  C.~N.~I.~S.~M. di Salerno,  Universit\`a
  degli Studi di Salerno, Via S. Allende, I-84081 Baronissi (Sa), Italy.}
\author{S.~Capponi} \affiliation{Laboratoire de
Physique Th\'eorique, CNRS-UMR5152, CNRS \& Universit\'e de
Toulouse, F-31062 Toulouse, France.}
\author{D.~Poilblanc} \affiliation{Laboratoire de
Physique Th\'eorique, CNRS-UMR5152, CNRS \& Universit\'e de
Toulouse, F-31062 Toulouse, France.}

\pacs{{75.40.Cx},{75.10.Jm}}
\date{\today}
\begin{abstract}
Recent experiments have revealed the role of staggered
Dzialoshinskii-Moriya interaction in the magnetized phase of an 
antiferromagnetic spin 1/2 two-leg
ladder compound under a uniform magnetic field. We
derive a low energy effective field theory describing
a magnetized two-leg ladder with a weak staggered Dzialoshinskii-Moriya
interaction. This theory predicts the persistence of the spin gap
in the magnetized phase, in contrast to standard two-leg ladders,
and the presence of bound states in the excitation spectrum. Such
bound states are observable in Raman scattering measurements. These
results are then extended to intermediate Dzialoshinskii-Moriya interaction
using Exact Diagonalizations.
\end{abstract}
\maketitle

Antiferromagnetic systems with spin gaps have been the object of
great theoretical and experimental interest for the last two
decades. In particular, ladder
systems\cite{dagotto_2ch_review,dagotto_supra_ladder_review} have
been investigated first theoretically as toy models for
high-temperature superconductivity\cite{rice_srcuo}, and then
experimentally.\cite{takano_spingap} Ladder systems with an even
number of legs are known to possess a spin gap, while ladders with
an odd number of legs are gapless, a result reminiscent of the one
of Haldane for antiferromagnetic spin-S chains.\cite{haldane_gap}
Upon application of a magnetic field, the gap of two-leg ladder
systems can be closed, leading to the formation of a Luttinger
liquid
state.\cite{chitra_spinchains_field,furusaki_dynamical_ladder} In
such a state, there exists a quasi-long range antiferromagnetic
order, and inter-ladder exchange leads, at low enough temperature, 
to the formation of a N\'eel
state.\cite{giamarchi_coupled_ladders} Such field induced magnetic
ordering is the quasi-one dimensional analogue of magnon Bose
condensation.\cite{nikuni00_tlcucl3_bec}

Recent Nuclear Magnetic
Resonance (NMR) experiments\cite{clemancey06_cuhpcl} have revealed that the
compound\cite{chiari_cuhpcl}
Cu$_2$(C$_5$H$_{12}$N$_2$)$_2$Cl$_4$ (Copper diazacycloheptane to be
abbreviated CuHpCl), originally believed to be a simple spin ladder
material\cite{chaboussant_cuhpcl,chaboussant_nmr_ladder,chaboussant_mapping}, 
possesses staggered
Dzialoshinskii-Moriya\cite{dzyalo_interaction,moryia_asym_int} (DM)
interactions along the rungs. Although recent
Inelastic Neutron Scattering experiments suggest that this
material is closer to a three-dimensional coupled dimer
system~\cite{stone02_cuhpcl} than to a real ladder, it was shown
theoretically (using a ladder description) that including such a
spin anisotropy greatly improves the low temperature fits of the
magnetization curve~\cite{miyahara06_dm,capponi06_ladder_dm} and of the
anomalous behavior of the specific heat in magnetic
field.~\cite{capponi06_ladder_dm} Due to the presence of these
interactions, the spin gap persists in the magnetized phase, and a
staggered magnetization is present even above the temperature for
antiferromagnetic ordering.\cite{clemancey06_cuhpcl}
The outline of the paper is the following. In
Sec.~\ref{sec:introduction}, some basic results on two-leg
ladders with staggered DM interactions are recalled.
Then, in Sec.~\ref{sec:boso},
we develop the field theoretical approach and compute the Raman
scattering spectra in the Fleury-Loudon
approximation\cite{fleury_raman,moriya_raman}. We obtain
 well defined peaks below a continuum.   The predictions of
 the field theoretical treatment are then compared to Exact
 Diagonalization (ED) results in
 Sec.~\ref{sec:numer}. The existence of peaks is confirmed, and is
 shown to persist to intermediate couplings.

\section{Introduction} \label{sec:introduction}

The Hamiltonian of the spin 1/2 Heisenberg two-leg ladder reads:
\begin{eqnarray}
  \label{eq:ladder}
  H&=&\sum_{n,p=1,2} \!\! J_\parallel \mathbf{S}_{n,p} \cdot
  \mathbf{S}_{n+1,p} + \sum_n J_\perp  \mathbf{S}_{n,1} \cdot
  \mathbf{S}_{n,2} \nonumber\\
& & - h \sum_{n,p=1,2} S_{n,p}^z,
\end{eqnarray}
where $\mathbf{S}_{n,p}$ is a spin 1/2 operator, $J_\perp,J_\parallel >0$
and $h=g\mu_B H$.
The staggered DM interaction is:
\begin{eqnarray}\label{eq:dm-stag}
  H_{DM}=\sum_n (-)^n \mathbf{D}\cdot (\mathbf{S}_{n,1}\times \mathbf{S}_{n,2})
\end{eqnarray}
with $\mathbf{D}=D\hat{y}$.

The point group of  the Hamiltonian
Eqs.~(\ref{eq:ladder})--~(\ref{eq:dm-stag}) is
$C_{2v}$.\cite{miyahara06_dm}
 It includes the
identity, the reflection symmetry w.r.t. a rung (or ``parity''),
the inversion symmetry w.r.t the center of a plaquette and the
combined operation of the two last ones.
The Hamiltonian~(\ref{eq:ladder})--(\ref{eq:dm-stag}) is also
invariant under translation of two lattice spacing $n\to n+2$.

\section{Bosonization study} \label{sec:boso}
We will derive a weak
coupling low-energy theory applying bosonization
techniques\cite{giamarchi_book_1d,gogolin_1dbook,nagaosa_book_sc}
valid for weak rung couplings. Then, in Sec.~\ref{sec:strong},
 we will consider the opposite
limit of strong rung exchange and derive an effective spin chain
Hamiltonian.\cite{chaboussant_mapping,mila_field} By bosonizing the
effective spin chain Hamiltonian in the limit of weak DM interaction,
we will show that the low energy theory derived for weak rung
couplings can be extended to strong rung exchange. In
Sec.~\ref{sec:raman}, we
analyze the effective field theory, and show that it supports
breather type excitations, in analogy with spin chains with staggered
DM interactions along the
chains.\cite{oshikawa_cu_benzoate_short,oshikawa_cu_benzoate}

\subsection{Weak coupling case}\label{sec:weak}

In this section, we derive an effective field theory describing the
low-energy, long-wavelength excitations of the Hamiltonian
Eq.~(\ref{eq:ladder})--~(\ref{eq:dm-stag}), valid in the limit
$J_\perp, D \ll J_\parallel$.
The bosonized representation of the Hamiltonian~(\ref{eq:ladder}) has
been derived for $J_\perp \ll J_\parallel$  in
Refs.~\onlinecite{strong_spinchains,strong_spinchains_long,shelton_spin_ladders,orignac_2spinchains,chitra_spinchains_field}.
We follow the notations of Ref.~\onlinecite{orignac_2spinchains}.
Using the bosonized representation for the spin operators ($p=1,2$):
\begin{eqnarray} \label{bosonized-spin}
S_p^{+}(x) & = &\frac{e^{-\imath \theta_p(x)}}{\sqrt{2\pi a}}
\left[e^{-\imath \frac{\pi x} a}+\cos 2\phi_p(x) \right],\nonumber \\
S_p^z(x) & = & -\frac{1}{\pi}\partial_{x}\phi_p +
e^{\imath \frac{\pi x} a} \frac{ \cos 2\phi_p(x)}{\pi a},
\end{eqnarray}
\noindent where $a$ is a lattice cutoff, and
$[\phi_p(x),\partial_x \theta_{p'}(x')]=i\pi\delta_{p,p'}\delta(x-x')$,
the Hamiltonian of the two-leg ladder is expressed in terms of the
fields
$\phi_{a}=\frac{\phi_{1}-\phi_{2}}{\sqrt{2}}$ and
$\phi_{s}=\frac{\phi_{1}+\phi_{2}}{\sqrt{2}}$ giving:
\begin{widetext}
\begin{eqnarray}
H & = & H_s +H_a \nonumber\\
H_s & = &\int \frac{dx}{2\pi}\left[ u_s K_s (\pi \Pi_s)^2
+\frac{u_s}{K_s}(\partial_{x}\phi_s)^2
\right] -\frac h \pi \int dx \partial_x \phi_s +\frac{2 J_\perp a}{(2\pi\alpha)^2}\int dx\cos(\sqrt{8}\phi_s),
\label{XXZ-final} \\
H_a & =&  \int \frac{dx}{2\pi}\left[ u_a K_a (\pi \Pi_a)^2
+\frac{u_a}{K_a}(\partial_{x}\phi_a)^2
\right] +\frac{2J_\perp a}{(2\pi\alpha)^2}\int dx\cos(\sqrt{8}\phi_a)
+\frac{2 J_\perp a}{(2\pi\alpha)^2}\int dx\cos(\sqrt{2}\theta_a),
\nonumber
\end{eqnarray}
\end{widetext}
\noindent where we have written $\partial_x \theta_p =\pi \Pi_p$.
It can be shown\cite{strong_spinchains,orignac_2spinchains} that for
$h=0$ the
Hamiltonian (\ref{XXZ-final}) has gapped excitations, with a gap $\Delta$ of
order $J_\perp$ and that the fields have the following expectation
values $\langle\phi_s\rangle =\pi/\sqrt{8}$
and $\langle \theta_a \rangle = \pi/\sqrt{2}$. Using
(\ref{bosonized-spin}), we find the bosonized form of $ H_{DM}$:
\begin{widetext}
\begin{eqnarray}
  \label{eq:dm-bosonise}
   H_{DM}=D \int dx \left[\frac{1}{\pi\sqrt{2\pi a}} (\partial_x \phi_2
   \cos \theta_1 - \partial_x \phi_1
   \cos \theta_2) - \frac{1}{\pi a \sqrt{2\pi a}} (\cos \theta_1 -\cos
   \theta_2) \cos 2\phi_1 \cos 2\phi_2\right].
\end{eqnarray}
\end{widetext}
In the presence of a nonzero magnetic
field the  term $-h
\int dx \partial_x \phi_s/\pi$ has non-trivial effects on the
spectrum.\cite{chitra_spinchains_field} For $D=0$, it induces a
commensurate-incommensurate
transition for
$h=\Delta$.\cite{japaridze_cic_transition,pokrovsky_talapov_prl} In
the commensurate state ($h < \Delta$), the system remains gapped and
non magnetized, whereas in
the incommensurate phase ($h>\Delta$) it is gapless and has a uniform
magnetization $m>0$.
If a small $D\ne 0$ is turned on  in the incommensurate phase,
the existence of a nonzero magnetization allows the substitution
 $\langle \partial_x
\phi_{1,2} \rangle \to
m$ in Eq.~(\ref{eq:dm-bosonise}) , giving a
term $m  D (\cos \theta_1 -\cos
\theta_2)$. Taking into account the fact that, in the incommensurate
phase, one still has $\langle \theta_a\rangle =\pi/\sqrt{2}$, this
term reduces to $ m D \langle \sin \theta_a/\sqrt{2}\rangle \sin
(\theta_s/\sqrt{2})$. The gap in the antisymmetric modes is given
by\cite{orignac_2spinchains}: $\Delta_{DM} \sim J_\parallel
(J_\perp/J_\parallel)^{\frac{2K_a}{4K_a-1}}$, giving$\langle \sin
\theta_a/\sqrt{2}\rangle \sim (J_\perp/J_\parallel)^{1/(16K_a-4)}$ For a magnetic field not
too strong, $K_a\simeq 1/2$ so that $\langle \sin
\theta_a/\sqrt{2} \rangle \sim (J_\perp/J_\parallel)^{1/4}$.

The low energy Hamiltonian describing the behavior of
a magnetized two-leg ladder with staggered
Dzialoshinskii-Moriya interaction is
therefore:
\begin{eqnarray}\label{eq:sinegordon-weak}
  H_s&=&\int \frac{dx}{2\pi} \left[ u_s K_s (\pi \Pi_s)^2 +\frac{ u_s}{
  K_s} (\partial_x \phi_s)^2\right]\nonumber \\ &&
+
\int dx   \frac{ D m \langle \sin (\theta_a/\sqrt{2})\rangle}{2\pi}  \sin \frac{\theta_s}{\sqrt{2}}.
\end{eqnarray}
Such a description is valid for energy scales much lower than
$|h-\Delta|$, and is therefore only applicable in a system which
is sufficiently magnetized. Let us briefly discuss its symmetries.
It is obvious that the Hamiltonian~(\ref{eq:sinegordon-weak}) is
invariant under parity and translation by two sites.
From~(\ref{bosonized-spin}), translation by one lattice site
amounts to $\theta_{1,2} \to  \theta_{1,2}+\pi$ and $\phi_{1,2}
\to  \phi_{1,2}+\pi/2-\pi m a $, and interchange of the chains to
$\theta_1 \leftrightarrow \theta_2$ and $\phi_1 \leftrightarrow
\phi_2$. The combination of these two transformations leaves the
Hamiltonian invariant. Such combination amounts to $\theta_s \to
\theta_s + \pi \sqrt{2}$, $\theta_a \to -\theta_a$ and $\phi_s \to
\phi_s  + \pi/\sqrt{2}(1-2m a)$ and $\phi_a \to -\phi_a$. It is
immediately seen that this combined transformation leaves the
Hamiltonian~(\ref{eq:sinegordon-weak}) invariant. The
Hamiltonian~(\ref{eq:sinegordon-weak})  is the one of a
sine-Gordon model, and its spectrum is massive provided
$K_s>1/16$. Since in the magnetized
phase\cite{chitra_spinchains_field}  $K_s\simeq 1/2$, the ladder
with staggered DM interaction remains gapped for all $h$, as found
in DMRG and ED.\cite{miyahara06_dm,capponi06_ladder_dm} The gap is
of order $\Delta_{DM} \sim J_\parallel
(Dm/J_\parallel)^{8K_s/(16K_s-1)} \sim
J_\parallel(mD/J_\parallel)^{4/7}$. It can be seen from
Eq.~(\ref{bosonized-spin}) that this implies a staggered
magnetization with: $ \langle S_{n,1}^x\rangle= -\langle
S_{n,2}^x\rangle \sim J_\parallel (Dm/J_\parallel)^{1/(16K_s-1)}
\sim J_\parallel(mD/J_\parallel)^{1/7}$. Since the dual field of
$\phi_s$ is ordered, the correlation functions of the non-uniform
magnetization along $z$ present an exponential decay instead of
the power law decay obtained for
$D=0$.\cite{furusaki_dynamical_ladder,giamarchi_coupled_ladders}

\subsection{Strong coupling case}
\label{sec:strong}

In this section, we consider again the two-leg ladder,
Eqs.~(\ref{eq:ladder})--~(\ref{eq:dm-stag}), but this time,
we assume a strong coupling limit such that $J_\perp \gg J,D$. In that
case, it is reasonable to diagonalize first the rung interaction and
derive an effective spin chain model to describe the behavior of the
system under
field.\cite{chaboussant_mapping,mila_ladder_strongcoupling}
We can write:
\begin{eqnarray}
  \label{eq:strong}
  S_{n,1}^z =S_{n,2}^z&=&\frac 1 4 (1+ 2 \tau_n^z) \nonumber \\
  S_{n,1}^+ =-S_{n,2}^+&=&-\frac 1 {\sqrt{2}} \tau_n^+
\end{eqnarray}
where the pseudospin $\tau_n^z=-1/2$ when the $n-$th rung is in the singlet
state, and  $\tau_n^z=+1/2$ when the $n-$th rung is in the triplet
state with $S_{n,1}^z +S_{n,2}^z=1$. In the absence of the DM
interaction, the Hamiltonian of the effective spin 1/2 chain
reads\cite{mila_ladder_strongcoupling,giamarchi_coupled_ladders}:
\begin{eqnarray}
  \label{eq:spinchain}
  H&=&\frac{J_\parallel}{2} \sum_n (\tau_n^+ \tau_{n+1}^- + \tau_n^-
  \tau_{n+1}^+ + \tau_n^z \tau_{n+1}^z) \nonumber\\
&-&(h-J_\perp-J_\parallel)  \sum_n \tau_n^z
\end{eqnarray}
Deriving a strong coupling expansion for Eq.~(\ref{eq:dm-stag}), we find that:
\begin{eqnarray}
  \label{eq:dm-strong}
  H_{DM}=\frac{D}{\sqrt{2}} \sum_n (-1)^n \tau_n^x.
\end{eqnarray}
Note that substituting simply (\ref{eq:strong}) in (\ref{eq:dm-stag})
yields a result which is incorrect by a factor 1/2. The reason for
this is that the operators $S_{n,1/2}^z$ do not annihilate the singlet
state contrarily to their low energy representation in terms of the
pseudospins. As a result of this, some of the contributions to
(\ref{eq:dm-strong}) are missed if we simply replace in
(\ref{eq:dm-stag}) the spin operators by their strong coupling
expansion~(\ref{eq:strong}).

Let us briefly review the symmetries of the strong coupling model.
First, we notice that the model is invariant under a translation of
two lattice spacing $n \to n+2$.
Second, it is also invariant under inversion $n \to -n$. Third, it is
also invariant under a reflexion around the middle of a bond $n \to
1-n$ (or a translation of one lattice site)
 combined with a rotation of $\pi$ around the $z$ axis.
If we turn to the two-leg ladder, we notice that
that this latter symmetry was apparently
not present in the original model. It is
in fact the inversion symmetry around the center of a plaquette in
disguised form.
Under inversion, singlet states change
sign (as $|\uparrow_1 \downarrow_2\rangle -|\uparrow_2
\downarrow_1\rangle \to -|\uparrow_2
\downarrow_1\rangle - |\uparrow_1 \downarrow_2\rangle$),
but triplet states do not. As a result, operators that transform
singlet state into triplet one are odd under inversion, and
operators diagonal on singlet or triplet states are even.
Thus, in the effective model,  inversion becomes the transformation
$\tau_n^+ \to -\tau_{1-n}^+$ and   $\tau_n^z \to +\tau_{1-n}^z$,
i. e. a reflection around the middle of a bond combined with a
rotation of $\pi$ around the $z$ axis.

Eq.~(\ref{eq:dm-strong}) shows that  the ladder under strong field
with a small staggered Dzialoshinskii-Moriya interaction
 becomes equivalent to a XXZ spin chain in a
staggered magnetic field. In the limit $D\ll J_\perp$ ,
this problem can be treated using
bosonization\cite{luther_chaine_xyz,giamarchi_book_1d},
and the amplitude of the
perturbation can even be found
exactly.\cite{lukyanov_ampl_xxz,lukyanov_spinchain_asymptotics,lukyanov_xxz_asymptotics}
Similar results are known to hold in the case of spin 1/2 chains with
a longitudinal staggered Dzialoshinskii-Moriya
interaction.\cite{oshikawa_cu_benzoate,essler99_cu_benzoate,essler_benzocu} Our bosonized theory reads:
\begin{eqnarray}
  \label{eq:boso-strong}
  H=\!\int \!\frac{dx}{2\pi} \left[ uK (\pi \Pi)^2\! +\! \frac u K (\partial_x
    \phi)^2\right] \!+\!\frac{\lambda D}{ a \sqrt{2}}\int\!\!dx \cos \theta,
\end{eqnarray}
where $u$ and $K$ can be obtained from (\ref{eq:spinchain}) by Bethe
Ansatz techniques\cite{giamarchi_coupled_ladders} provided that $D
\ll J_\parallel$. This is a quantum sine-Gordon (SG)
model.\cite{coleman_equivalence} For $K>1/8$, its spectrum
contains massive solitons~\cite{note_soliton}, which means that
the original Hamiltonian possesses a spin gap. The magnitude of
the gap will be $\Delta_{DM} \propto J_\parallel
(D/J_\parallel)^{4K/(8K-1)}$ provided that $K>1/8$. We note that
this strong coupling bosonized Hamiltonian is of the same form as
the one derived in Sec.~\ref{sec:weak}, indicating that the weak
and the strong coupling regime are continuously connected. The
formal correspondence between the two Hamiltonians is obtained by
the canonical transformation $\theta=\frac \pi 2 -
\theta_s/\sqrt{2}$, $\phi=-\phi_s\sqrt{2}$, $K=2K_s$. In terms of
the sine-Gordon model~(\ref{eq:boso-strong}), the symmetries of
the lattice Hamiltonian become continuous translational
invariance, invariance under the transformation $\theta(x) \to
\theta (-x)$ and $\phi(x) \to -\phi (-x)$ and $\theta \to
\theta+\pi$, $D \to -D$. This latter symmetry is a $Z_2$ gauge
symmetry resulting from the impossibility of discerning odd and
even sites in the continuum limit. A final symmetry of the
bosonized Hamiltonian is $\theta \to -\theta$ and $\phi \to
-\phi$. The latter symmetry is a consequence of the linear
dispersion of excitations for $D=0$ in the scaling limit and is
absent in the lattice system. From the semiclassical
treatment\cite{dashen_sinegordon}, it is seen that parity
 transforms solitons into
antisolitons and vice-versa while leaving the breathers invariants.
Indeed, classically, the soliton (resp. antisoliton)
of the Sine-Gordon model is given by $\theta(x)=
4\arctan{(\exp{(x/l)})}$ (resp. $\theta(x)=
4\arctan{(\exp{(-x/l)})}$) so that the two are exchanged by a parity
transformation. States odd or even under parity can be constructed
by  linear
combination of soliton and antisoliton states. Lastly, we note that the bound
states corresponding to the SG breathers should be even under parity as
breathers contain an equal number of solitons and
antisolitons~\cite{note_breathers}.

An interesting aspect of the Hamiltonian~(\ref{eq:boso-strong}) is
that it can possess breathers (bound states of solitons) as excited
states provided $K>1/4$.\cite{solyom_revue_1d,dashen_sinegordon,luther_chaine_xyz} It can be seen from 
Ref.~\onlinecite{giamarchi_coupled_ladders} that this condition is always
satisfied. If $M_0\propto \Delta_{DM}$
is the mass of a soliton, the masses of the
breathers will be given by:
\begin{eqnarray}
  \label{eq:breather-masses}
  M_n = 2M_0 \sin \left(\frac \pi 2 \frac n {8K -1}\right),
\end{eqnarray}
with $n$ integer and $1\le n < 8K-1$. Under the transformation $\phi
\to -\phi$ and $\theta \to -\theta$ breathers with an odd $n$ are odd,
while breathers with an even $n$ are even.

In the case of a spin chain with a longitudinal staggered
Dzialoshinskii-Moriya interaction, the effective theory is also a
sine-Gordon model as Eq.~(\ref{eq:boso-strong}), but with $K=1/2$
for not too large applied magnetic
field. This leads to a smaller number of breathers than in the ladder case.\cite{asano_breather,zvyagin05_esr_cu_pyrimidine,bertaina_bacu2ge2o7_esr,nojiri_breather}  Such breathers
have been be detected in Electron Spin Resonance (ESR) 
measurements.\cite{asano_breather,feyerherm_breather,wolter03_field_gap,bertaina_bacu2ge2o7_esr,kenzelmann_breather,zvyagin_breather,wolter_canting,zvyagin05_esr_cu_pyrimidine,morisaki_dzialmo,nojiri_breather,morisaki07_staggered_dm}
In the case of the ladder system however, ESR measurements cannot be
straightforwardly interpreted. Indeed, ESR absorption intensity is
proportional to an autocorrelation function of $(S_1^x + S_2^x)$, but
this operator is vanishing in the strong coupling limit according to
(\ref{eq:strong}) and in the weak coupling limit it is mixing the
symmetric and the antisymmetric modes, thus complicating the
interpretation of the spectra. We suggest in Sec.~\ref{sec:raman} 
that Raman scattering intensity measurements give a direct access to
these breather modes in the ladder systems.  

Let us now discuss the variation of the number of breathers with the
magnetization in the case of the ladder. For a magnetization near
zero, we have $K\alt 1$ so we expect to find 6 breather excitations
with masses given by (\ref{eq:breather-masses}). As the magnetization
increases, $K$ decreases, and for $K<7/8$ there will be only 5
breathers. When $K=3/4$, i.e. at half the saturation magnetization,
there are only 4 breathers. As the magnetization increases beyond this
value, $K$ becomes an increasing function of the magnetization, and
the number of breathers becomes again increasing.

When the magnetization is at half the saturation value, a more
quantitative analysis of the gap becomes possible.
Indeed, in that case $\langle \tau_n^z\rangle=0$ and 
the exponent $K$ and the velocity $u$ are given as:\cite{luther_chaine_xxz,haldane_xxzchain}
\begin{eqnarray}\label{eq:exp-half-sat}
  K&=&3/4, \\
u&=& \frac{3 \sqrt{3}} 8 J_\parallel a. \nonumber
\end{eqnarray}
Moreover, the amplitude $\lambda$ is given by the integral:\cite{lukyanov_ampl_xxz,lukyanov_xxz_asymptotics,hikihara03_amplitude_xxz}
\begin{widetext}
\begin{eqnarray}\label{eq:amplitude}
  \lambda^2=\frac 1 {4\left(1-\frac 1 {2K}\right)^2}
  \left[\frac{\Gamma\left(\frac{1/2}{2K-1}\right)}{2\sqrt{\pi}
      \Gamma\left(\frac K {2K-1}\right)}\right]^{\frac 1 {2K}}
  \exp\left[-\int_0^\infty \frac {dt}{t} \left(\frac{\sinh\left(\frac t
        {2K}\right) }{\sinh t \cosh\left[\left(1-\frac 1 {2K}\right)
        t\right]} -\frac{e^{-2t}}{2K}\right)\right],
\end{eqnarray}
\end{widetext}
\noindent which can be evaluated using
Ref.~\onlinecite{abramowitz_math_functions}, Eq. (6.3.22) giving:
\begin{eqnarray}
  \lambda^2=\frac{9}{4 \pi^{2/3}}
  \left(\frac{\Gamma(2/3)}{\Gamma(1/3)}\right)^2.
\end{eqnarray}
This expression yields $\lambda\simeq 0.51769$ in agreement with Table I in
Ref.~\onlinecite{hikihara03_amplitude_xxz}.
Finally, using the expression of the gap of the sine Gordon model from 
Ref.~\onlinecite{lukyanov_sinegordon_correlations},  we obtain the quantitative
expression of the spin gap:
\begin{eqnarray}
  \label{eq:gap}
  \frac{\Delta_{DM}}{J_\parallel} = \frac{3 \sqrt{3}}{4 \sqrt{\pi}}
  \frac{\Gamma(1/10)}{\Gamma(3/5)} \left(\frac{\pi^{2/3}
      \Gamma(5/6)\Gamma(2/3)} {\Gamma(1/6) \Gamma(1/3)}  \sqrt{\frac 2
      3}
    \frac D {J_\parallel} \right)^{3/5}
\end{eqnarray}
\noindent \emph{i.e.} $\Delta_{DM}/J_\parallel \simeq 1.67
(D/J_\parallel)^{3/5}$. Inserting the
result~(\ref{eq:exp-half-sat})  in Eq.~(\ref{eq:breather-masses})
leads to the prediction of four breathers with energies
$E_n=2\Delta_{DM} \sin(n \pi/10)$ ($n=1,\ldots,4$). We notice that
the energy of the lowest breather is smaller than the energy of
the soliton. If we assume that the spin gap coincides with the
energy of the lowest breather mode, we will find a spin gap
$E_1=1.032 J_\parallel (D/J_\parallel)^{3/5}$. This formula is in rough
agreement with the numerical data of Sec.~\ref{sec:numer}. To
obtain quantitative estimates of the spin gap away from half the
saturation magnetization, where (\ref{eq:amplitude}) is not
applicable anymore, one can use the numerical estimations of the
amplitude $\lambda$ obtained in Ref.~\onlinecite{hikihara03_amplitude_xxz}.
This will not be attempted here.

\subsection{Raman scattering intensity}
\label{sec:raman}

In that section, we determine the Raman scattering intensity of the
two-leg ladder under a magnetic field in the limit $D,J_\parallel\ll
J_\perp$. We will assume that the
Fleury-Loudon
approximation\cite{fleury_raman,moriya_raman} is
applicable, i.e. that the frequency of the radiation is much smaller
than the Mott gap of the material.\cite{shastry_raman}
 In that case, the (effective) Raman operator acts in the (restricted) Hilbert space
of the spin configurations and can be written as:
\begin{equation}\label{lf.eq}
\hat{O}_R= \sum_{\langle ij\rangle} J_{ij} (\hat{e}_{I}\cdot \hat{n}_{ij}) (\hat{e}_O\cdot \hat{n}_{ij}) {\bf S}_i \cdot
{\bf S}_j
\end{equation}
where the $\hat{n}_{ij}$ are unit vectors along the bond directions,
 $\hat{e}_I$  the polarization vector of incoming
light, and $\hat{e}_O$  the polarization vector of outgoing
light.
In principle, the strength of the Raman scattering $J_{ij}$ could depend on the geometry~\cite{freitas_raman}
but, as in most studies~\cite{natsume_ladder_raman}, we assume it to
be constant.

For simplicity,
here we restrict ourselves to simple geometrical setups, e.g. to the XX, YY
and X'Y' geometries
corresponding to $\hat{e}_{I}$ and $\hat{e}_O$, both along
the chains direction, both
perpendicular to the
chains and at 45-degrees from the crystal axes (X'=X+Y, Y'=X-Y), respectively.
The corresponding Raman operators can then be written as,
\begin{eqnarray}
\hat{O}_R^{XX} &=& \gamma \sum_{n} ({\bf S}_{n,1} \cdot {\bf S}_{n+1,1}
+ {\bf S}_{n,2} \cdot {\bf S}_{n+1,2})\, ,
  \label{eq:leg-op} \\
\hat{O}_R^{YY} &=& \gamma \sum_{n} {\bf S}_{n,1} \cdot {\bf S}_{n,2} \, ,
  \label{eq:rung-op}\\
\hat{O}_R^{X'Y'} &=&  \frac{1}{2}(\hat{O}_R^{XX} - \hat{O}_R^{YY}) \, .
\end{eqnarray}

In the absence of DM interaction,
it is
expected
that there is no Raman intensity for frequencies smaller than twice
the spin
gap.\cite{orignac_raman_short,bibikov_raman_ladder,donkov_raman_ladder,natsume_ladder_raman}
For frequencies equal to approximately twice the spin gap, a threshold
is obtained when interaction between magnons are
neglected\cite{orignac_raman_short,bibikov_raman_ladder} with the
Raman intensity behaving as $I_R(\omega) \sim
(\omega-2\Delta)^{1/2}$. When interactions are added, the threshold
becomes a peak.\cite{donkov_raman_ladder}
Turning to the ladder with Dzialoshinskii-Moriya interaction, in the
weak coupling case, the threshold  at twice the spin gap will be
obtained from the contribution of antisymmetric modes
$\theta_a/\phi_a$ to the Raman intensity.\cite{orignac_raman_short}
The symmetric modes will give extra peaks and thresholds but at energy
scales small compared to the gap. In the following, we will compute the Raman
intensity for frequencies smaller than  $2 J_\perp$. In that case, we
can use the strong coupling theory for the calculation. We will
consider two cases, that of a field parallel to the legs in
Sec.~\ref{sec:raman-leg}, and then that of a field parallel to the
rungs in Sec.~\ref{sec:raman-rung}.

\subsubsection{Electric field along the legs}
\label{sec:raman-leg}

In the strong coupling limit, we can use Eq.~(\ref{eq:strong}) to
rewrite the Raman operator (\ref{eq:leg-op}) as,
\begin{eqnarray}
  \label{eq:raman-strgcpl}
  \hat{O}_R^{XX}&=&\frac{\gamma}{2} \sum_n (\tau_n^+ \tau_{n+1}^-  +
  \tau_{n+1}^+\tau_n^- \nonumber\\
&+& (\tau_n^z+1/2)(\tau_{n+1}^z+1/2)),
\end{eqnarray}
yielding the expression:
\begin{eqnarray}
  \hat{O}_R^{XX}=\gamma\frac{H-H_{DM}+ (h-J_\perp-J_\parallel/2)  \sum \tau_n^z}{ J_\parallel} +\text{const.},
\end{eqnarray}
where $H$ is the full strong coupling effective spin chain
Hamiltonian. In the computation of the Raman scattering
intensity\cite{orignac_raman_short}, we had already noticed that the
time independent terms would drop out from the calculation. Therefore,
we can consider the effective Raman operator:
\begin{eqnarray}
   \hat{O}_R^{XX}=\gamma(h_{0} \sum_n \tau_n^z-H_{DM})/J_\parallel,
\end{eqnarray}
\noindent where $h_{\text{0}}=h-J_\perp-J_\parallel/2$.
Going to the continuum limit, it is convenient to introduce a
density for the Raman operator, $O^{XX}_R(x)$ so that
$\hat{O}_R^{XX} =\int O_R^{XX}(x)
dx$. Using the density of Raman operator, and the standard definition
of Raman scattering intensity, we finally find that:
\begin{eqnarray}
  I_{R,\parallel}(\omega)=\sum_n |\langle n,\mathbf{P}_n=0|O^{XX}_R(0)|0\rangle|^2
  \delta(\omega-E_n)
\end{eqnarray}
where $|0\rangle$ is the ground state (GS) of the system and
$|n,\mathbf{P}_n\rangle$ is the $n-$th excited state with
$\mathbf{P}_n$ its total momentum. The Raman operator being given by a
sum over all sites, it is easy to show by Fourier transformation that
Raman scattering is conserving photon momenta. This explains why only
the states with zero momenta (i.e. having the same momentum as the
ground state) contribute to the sum.

To proceed with the calculation, we need the bosonized expression of
the density of Raman operator. It is easy to show that it reads:
\begin{eqnarray}\label{eq:raman-boso-leg}
  O^{XX}_R(x)= \frac{\gamma}{J_\parallel} \left( \frac{\lambda D}{ a
    \sqrt{2}}  \cos \theta(x)  - \frac
    {h_{0}}  \pi \partial_x
  \phi\right)
\end{eqnarray}

The Raman intensity can now be found by applying the  Form Factor
method\cite{karowski_ff,mussardo_trieste_2001,controzzi_mott}.
 The form factors  of an operator are simply the matrix elements of this
operator between the ground state and an excited state.
In the sine-Gordon model,  the excited
states  are formed of solitons,
antisolitons and breathers with given rapidities.  Because in one
dimensional  integrable models collisions
between particles do not lead to production of new
particles\cite{dorey_smatrix_review}, the form factors are derived
from the sole knowledge of  the
exact S-matrices of the sine-Gordon
model.\cite{zamolodchikov79_smatrices,dorey_smatrix_review,mussardo_offcritical_review}

In our case, the relevant form factors have been already
computed by other authors in slightly different
contexts\cite{controzzi_mott,essler98_cu_benzoate}.
Using the symmetries of the operators in
Eq.~(\ref{eq:raman-boso-leg}), it is possible to further simplify the
expression of the Raman intensity.
 The sine-Gordon
Hamiltonian is invariant under the simultaneous transformations
 $\theta \to -\theta$ and $\phi \to -\phi$, and its eigenstates can
 be classified according to their parity under this transformation.
We note that the operator $\cos \theta$ is even under such
transformation, whereas the operator $\partial_x \phi$ is odd.
Therefore, the non-vanishing matrix elements of $\cos \theta$
are  between the ground state and
even states, whereas those of $\partial_x \phi$
are  between the ground state and the odd states.
An immediate consequence is that:
\begin{eqnarray}
  I_{R,\parallel}(\omega)=I_{R,\parallel}^{(o)}(\omega)+ I_{R,\parallel}^{(e)}(\omega),
\end{eqnarray}
where:
\begin{eqnarray}\label{eq:raman-xx-oe}
  I_{R,\parallel}^{(o)}(\omega)&=& \left(\frac{\gamma h_{0}}{\pi
        J_\parallel}\right)^2 \sum_{|n\rangle odd} |\langle n |\partial_x
  \phi |0\rangle|^2 \delta(\omega-E_n) \\
I_{R,\parallel}^{(e)}(\omega)&=& \left( \frac{\lambda \gamma D}{ a
   J_\parallel \sqrt{2}} \right)^2\sum_{|n\rangle even} |\langle n |\cos \theta |0\rangle|^2 \delta(\omega-E_n)\nonumber
\end{eqnarray}
The intensity $I_{R,\parallel}^{(e)}(\omega)$ has been computed
previously in Ref.~\cite{essler98_cu_benzoate} in the context of 
spin chains with staggered Dzialoshinskii-Moriya interactions. 
>From the results of Ref.~\onlinecite{essler98_cu_benzoate}, 
$I_{R,\parallel}^{(e)}(\omega)$ contains delta-function peaks  at
the frequencies of even breathers, and  thresholds at frequencies
equal to twice the soliton frequency, and to frequencies equal to the
sum of the masses of two breathers of identical parity. The lowest
threshold is therefore obtained at the frequency $\omega=2M_1$. We
notice that this threshold is below the energies of the breathers of
mass $M_3$ and $M_4$. Therefore, peaks are obtained in the continuum.
Of course, in a more realistic model, the conservation laws that make
the sine Gordon model integrable would be absent, and one should
expect that the peaks would be broadened by coupling with the
continuum. If the deviations of the lattice model from the continuum
sine-Gordon model are sufficiently important (for instance if the gap
is so large that the continuum limit is not justified), they may even
lead to the complete disappearance of the two peaks.

Turning to
the intensity $I_{R,\parallel}^{(o)}(\omega)$, we
use the equation of motion for the field $\theta$,
$\partial_t \theta=\frac u K   \partial_x \phi$,
to relate it  to the electrical conductivity of a
one dimensional Mott insulator.\cite{controzzi_mott,essler_mott_excitons1d}
In Ref.~\onlinecite{essler_mott_excitons1d}, it was shown that
if the sine-Gordon model describing the low energy charge excitations
of the Mott insulator possesses breathers in its spectrum, the
conductivity of the Mott insulator has delta peaks at the
breather frequencies. In the context of the Mott insulator, these peaks
were interpreted as exciton lines.\cite{essler_mott_excitons1d}
Translating the results of Ref.~\onlinecite{essler_mott_excitons1d}, we
conclude immediately that  $I_{R,\parallel}^{(o)}(\omega)$ possesses
 delta function peaks at
the frequencies of the odd breathers. Moreover, there are thresholds
at frequencies equal to twice the soliton mass and to the sum of two
breather masses of different parities. The first threshold in
$I_{R,\parallel}^{(o)}(\omega)$ is thus at the  frequency  $\omega=M_1+M_2$.

\begin{figure}[!ht]
  \includegraphics[width=9cm]{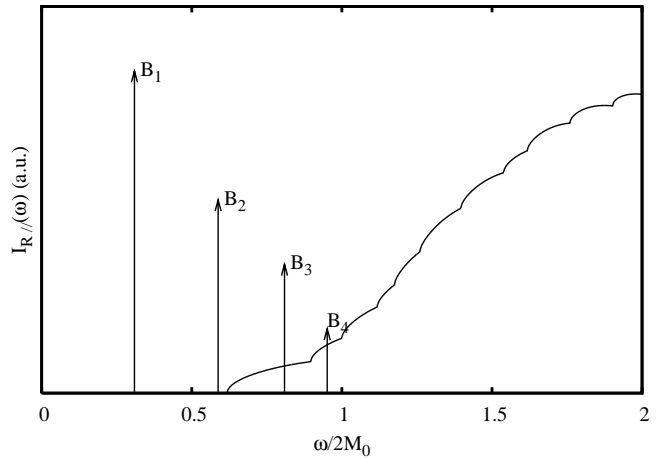}
  \caption{A sketch of the Raman intensity for polarization parallel
    to the legs. The arrows  represent the position of the
   the breathers $\delta$--peaks.
  The intensity in the peak, proportional to its height
  is decreasing with the breather index. The continuum starts at a
  frequency equal to twice the mass of the lightest breather. Above the threshold,
   resonances are expected instead of $\delta$-functions (beyond a SG analysis).}
  \label{fig:raman_par}
\end{figure}

Combining the two contributions, both the odd and the even
 breathers yield
peaks in the Raman intensity. Moreover, the Raman intensity exhibits
 thresholds at all
frequencies equal to the sum of two breather masses or to twice the
soliton mass. A qualitative
sketch of the Raman intensity is visible on the figure~\ref{fig:raman_par}.

In the special case of  $h_{\text{eff}}=0$,  the
intensity $I_{R,\parallel}^{(o)}$ is vanishing. Only the even breathers will
contribute to the Raman spectrum.

The weight of the delta peaks can be computed by the Form factor
expansion.\cite{babujian99_ff_sg_1,babujian02_ff_sg_2,lukyanov97_ff_sinegordon}
 From the general form factor expansion, we have that:
\begin{eqnarray}
  I_R(\omega) = \sum_n |\langle 0 |\hat{O}_R |B_n(0) \rangle |^2 \frac{\delta (\omega
  -E_n)}{E_n},
\end{eqnarray}
\noindent where $|B_n(0) \rangle$ is the state with one $B_n$ breather
of momentum zero, $E_n$ the energy of the breather and $\langle 0 |\hat{O}_R
|B_n(0) \rangle$ is the one breather form factor. Note that this formula
{\it a priori} applies to both $\parallel$ or $\perp$ setups
corresponding to the XX and YY (see below) polarizations.

Our task is thus to compute the one breather form factors for the two
operators $\partial_t \theta$ and $\cos \theta$. Let us first sketch
the idea behind the calculation.
The axioms of Form-Factor theory state that the existence of
breathers result in poles in the $n$-particle
form-factor (considered as complex function of the rapidities of
the particles) when the difference of two rapidities is equal to a
(purely imaginary) fusion angle.\cite{mussardo_offcritical_review,smirnov92_ff_book}
The residue at the pole is proportional to the $(n-2)$ particle, one
bound state form-factor. Therefore, the Form-factor for a single
soliton bound-state will be obtained from the residue of the two
particle form factor at the appropriate fusion angle.

To be more precise, in the case of the sine-Gordon model, one has
for the soliton-antisoliton form factor of an operator $O$:
\begin{eqnarray}
  F^O(\theta)_{\theta \to iu} =\frac{R}{\theta-i u} +\ldots
\end{eqnarray}
\noindent where:
\begin{eqnarray}
  R=\langle 0 | O | B \rangle  \frac{(-)^k i}{\sqrt{2}} \left|
    \mathrm{Res} S(\theta) \right|^{1/2},
\end{eqnarray}
with $=\langle 0 | O | B \rangle$ the form factor of the breather, $u$
the corresponding fusion angle, and $S$ the $S-$matrix for
soliton-antisoliton collision in the sine-Gordon model.
In the case of the sine-Gordon model\cite{babujian02_ff_sg_2},
the $k-$th fusion angle is given
by $u=\pi(1-k/\lambda)$, with $\lambda=8K-1$. The residue of the $S$-matrix is given by:
\begin{eqnarray}
   \left|
    \mathrm{Res} S(\theta) \right|_{\theta=i\pi(1-k/\lambda)}  =2
  \cot \left( \frac{\pi k}{2
      \lambda}\right) \prod_{\ell=1}^{k-1} \cot^2 \left( \frac{\pi \ell}{2
      \lambda}\right)
\end{eqnarray}

The soliton-antisoliton form-factor for $\partial_t \theta$ is given by:
\begin{eqnarray}
  \langle 0 |\partial_t \theta |\varphi_1,\varphi_2\rangle
  &=&\frac{4M}{\beta}
  \cos\!\left(\!\frac{ \varphi_1+\varphi_2}{2}\!\right)\nonumber\\
&&  \frac{\cosh\frac{1}{2}(i\pi -\varphi)}{\cosh\frac{\lambda}{2}(i\pi
    -\varphi)} F_{min}(\varphi),
\end{eqnarray}
\noindent where $\varphi=\varphi_1-\varphi_2$,
And the one of $e^{i\theta}$ by:
\begin{eqnarray}
  \langle 0 | e^{i \theta} |\varphi_1,\varphi_2\rangle
  = \frac{\cosh(\varphi/2) e^{\lambda(i\pi -\varphi)/2}}{\sinh
    \lambda({i\pi -\varphi})}  F_{min}(\varphi),
\end{eqnarray}

Where:

\begin{eqnarray}
  F_{min}(\varphi)&=& \exp \left[\int_0^\infty \frac{dt}{t} \frac{\sinh
      \frac t 2 \left( 1 - \frac 1 \lambda \right)}{\cosh \frac t 2
      \sinh \left(\frac t {2\lambda} \right) } \right. \nonumber\\
      & & \left. \frac{\sin^2 \frac{t}{2\pi} (\pi - i \varphi)}{\sinh t}\right]
\end{eqnarray}

As a function of the fusion angle, the energy of the breather is
$2M \cos(u/2)=2M \sin(k\pi/2\lambda)$. The Form factor of $\partial_t
\theta$ has poles only for odd $n$ whereas the Form factor of $e^{i
  \theta}$ has poles for both odd and even $k$. Hovever, the poles at
odd $k$ are the contribution of $\sin \theta$ and therefore, only the
poles for even $k$ must be considered in order to obtain the
contribution of the $\cos \theta$ term to the Raman intensity.
We find:
\begin{widetext}
\begin{eqnarray}
  |\langle 0 |\partial_ t \theta |B_{2n+1}(0)\rangle|^2  = C_1 \sin
  \left(\frac{\pi (2n+1)}{\lambda} \right) \prod_{\ell=1}^{2n} \tan^2
  \left(\frac{\pi \ell}{2 \lambda}\right) \exp\left[-2 \int_0^\infty \frac{dt}{t} \frac{\sinh
      \frac t 2 \left( 1 - \frac 1 \lambda \right)}{\cosh \frac t 2
      \sinh \left(\frac t {2\lambda} \right) }  \frac{\sinh^2
      \left(\frac{(2n+1)t}{2\lambda}\right)}{\sinh t}\right],
\end{eqnarray}
recovering the result in  Ref.~\onlinecite{essler_mott_excitons1d}, and:
\begin{eqnarray}
   |\langle 0 |\cos \theta |B_{2n}(0)\rangle|^2 = C_2 \sin^2
   \left(\frac{2\pi n}{\lambda} \right) \tan \left(\frac{\pi n}{\lambda}
 \right)  \prod_{\ell=1}^{2n-1} \tan^2
  \left(\frac{\pi \ell}{2 \lambda}\right) \exp\left[-2 \int_0^\infty \frac{dt}{t} \frac{\sinh
      \frac t 2 \left( 1 - \frac 1 \lambda \right)}{\cosh \frac t 2
      \sinh \left(\frac t {2\lambda} \right) }  \frac{\sinh^2
      \left(\frac{2n t}{2\lambda}\right)}{\sinh t}\right],
\end{eqnarray}
\end{widetext}
\noindent where $C_1$ and $C_2$ are independent of $n$. In the case of $K=3/4$,
we have 4 breathers.  The weight of the peak associated with
the first breather is obtained numerically as  $0.345 C_1$. For the peak
associated with the third breather the weight  is only  $0.0116C_1$
i.e. about 3\% of the weight of the first
breather. Similarly, the weight of  the second breather, we
 is $0.0367 C_2$ and the weight of the
fourth breather is $0.0058C_2$ i. e. about 16\% of the intensity
associated of  the second breather. Note that the ratio $C_2/C_1$
depends not only on the Luttinger parameter but also on the ratio
$D/J_\parallel$.

\subsubsection{Electric field along the rungs}
\label{sec:raman-rung}

In the case of an electric field parallel to the rungs, the Raman
operator (\ref{eq:rung-op}) can be rewritten,
in the strong coupling limit, as:
\begin{eqnarray}
  \label{eq:rung-op-sc}
  \hat{O}^{YY}_{R}=\gamma \sum_n \left(\tau_n - \frac 1 4 \right) \, .
\end{eqnarray}
In the absence of the Dzialoshinskii-Moriya interaction, this term is
proportional to the total magnetization, which leads to a vanishing
Raman intensity for $\omega \ne 0$. However, this result is only valid
for frequencies $\omega \ll J_\perp$ since the strong coupling
approximation eliminates the states $S^z=0,1$ that are placed at
energies $\sim J_\perp$ above the triplet and the singlet. From the results of
Ref.~\onlinecite{orignac_raman_short}, we know that for $\omega$ equal to twice
the spin gap, a photon can create a pair of magnons, resulting in a
non-zero Raman intensity. Therefore, in a
two-leg ladder with $D=0$ the Raman intensity should be zero in an
interval $[0,2(J_\perp-J_\parallel)]$.

When we turn on $D$, the total magnetization is not anymore a good
quantum number, and we expect a non-zero Raman intensity also
for energies small compared to $J_\perp$. Using bosonization, we can
express the Raman intensity as:
\begin{eqnarray}
  I_{R,\perp}(\omega) = \frac{K^2 \gamma^2}{\pi^2 u^2} \sum_n E_n^2 |\langle 0
  |\theta |n \rangle|^2 \delta(P_n) \delta(\omega -E_n)
\end{eqnarray}

\begin{figure}[!ht]
  \includegraphics[width=9cm]{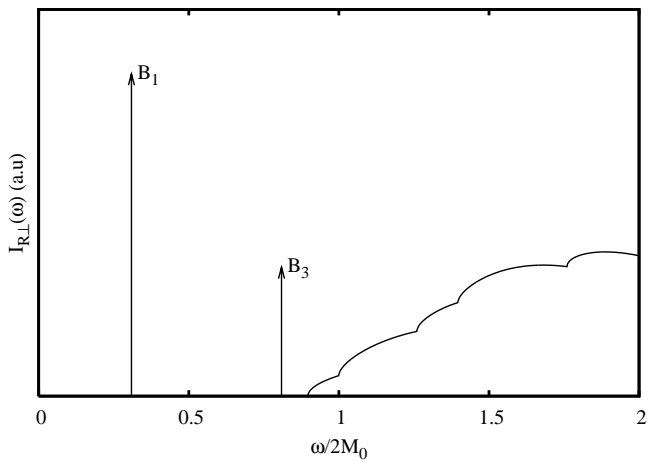}
  \caption{A sketch of the Raman intensity for polarization parallel
    to the rungs. The arrows represent the position of the
   the breathers $\delta$--peaks.
The intensity in the peak, proportional to its height
  is decreasing with the breather index. In contrast to the case of
   polarization along the legs, only odd breathers contribute to the
   Raman intensity.
  The continuum starts at a
  frequency equal to the sum of the masses of the lightest odd and
  even breathers, which is higher than in the case of polarization
  along the legs. }
  \label{fig:raman_perp}
\end{figure}

It is obvious that $ I_{R,\perp}(\omega) \propto
I^{(o)}_{R,\parallel}(\omega)$ and therefore only contains
contributions from odd states.  Moreover, the continuum starts at the
frequency $M_1+M_2$ which is larger than in the case of a polarization
along the legs. The behavior of  $ I_{R,\perp}(\omega)$ is sketched on
Fig.~\ref{fig:raman_perp}. Again, as we noted in
Sec.~\ref{sec:raman-leg}, the breather $B_3$ has a higher energy than
the lowest two-$B_1$ breather excitation. So the breather $B_3$ may
be unstable in the more realistic case, and may appear as a broad
resonance or be completely absent.

In the case of $X'Y'$ polarization, it is obvious that the
resulting Raman operator being a linear combination of $O_R^{XX}$
and $O_R^{YY}$~\cite{note_XY}, the resulting intensity will be
qualitatively similar to $I_{R,\parallel}$ shown on Fig.~\ref{fig:raman_par}.

\section{Numerical Exact Diagonalization Results}\label{sec:numer}

\subsection{Introduction}

Since, strictly speaking, the previous analytical approaches are only
valid in the limit $D\ll J_\parallel\ll J_\perp$, it is of great
interest to benefit also
from a complementary numerical approach valid within a
broader parameter range.
In this Section, coming back to the original ladder
model of Eq.~(\ref{eq:ladder})--(\ref{eq:dm-stag})
(again with alternating DM vector and
external magnetic field perpendicular to it), we perform
extensive Exact Diagonalizations (ED) on finite periodic $2\times L$ clusters
(the length $L$
ranging from 4 to 14) using Lanczos (for GS static and dynamical properties)
or Davidson (for computing low-energy spectra) algorithms.

To illustrate our results, we use a typical parameter
$J_\parallel=0.2 J_\perp$ consistent with a (crude) theoretical
description of the molecular solid
CuHpCl~\cite{chaboussant_nmr_ladder}. When not specified
otherwise, the largest parameter $J_\perp$ sets the energy scale.
Fortunately, for such a small $J_\parallel/J_\perp$ ratio, finite
size effects remain small and results are quite reliable, in
particular for $D>0.1$. Note that finite size scaling can also
often be performed to improve further the accuracy of the
computation. Interestingly, accuracy is getting better for
increasing $D$ (at finite fields) i.e. precisely in the range of
parameters for which the analytic treatment is becoming
unreliable. The bosonization and ED techniques are therefore
complementary. Note also that, despite the lack of numerical
accuracy in that limit, we have found that the ED results are
consistent with the bosonization when $D\rightarrow 0$, as
discussed later.

\begin{figure}[!ht]
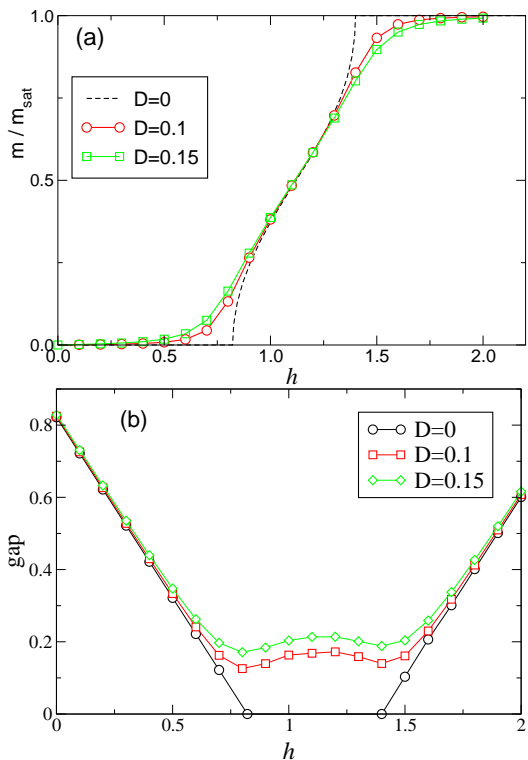

\includegraphics[width=0.8\linewidth]{fig3a}
\includegraphics[width=0.8\linewidth]{fig3b}
\caption{(Color online) Magnetization (a) and gap (b) vs (reduced) magnetic
field for finite $D=0.1$ and $D=0.15$, compared to
the $D=0$ case, computed on a $2\times 12$ ladder.
Note that the $h=h_{c,1}$ singularity at $D=0$ disappears for finite $D$.
(The behavior of the gap at $D=0$ is schematic). All energies are in
units of $J_\perp$.}
\label{fig:magnet}
\end{figure}

As mentioned in previous studies~\cite{miyahara06_dm,capponi06_ladder_dm},
the DM term has been shown to produce a smooth magnetization
curve, even for finite clusters, and a gap above $h_{c1}$.
For convenience we reproduce here
in Fig.~\ref{fig:magnet}(a) and (b) the (reduced) magnetization curve
and the gap, respectively, as a
function of magnetic field for $D=0.1$ and $D=0.15$, values of the
DM interaction that we shall consider later on, and also for
$D=0$, for comparison.

In the following, for simplicity, we shall
focus on the special field such that $m= 0.5 m_{sat}$. From direct
inspection of the magnetization curve, one gets $h \sim 1.1
J_\perp$~\cite{note_h}, almost independently of the value of $D$
since all curves seem to cross at his point. Note that, for such a
field, bosonization predicts $K=3/4$ and four breathers.

\subsection{Low energy spectra: solitons and bound states}

Prior to the calculation of the Raman spectrum, it is interesting
and useful to first explore the structure of the low energy
spectrum. The clusters being periodic (ring geometry), one can
take full advantage of the lattice translation symmetry (the unit
cell contains two rungs) so that eigenstates are labelled
according to their momentum along the legs. We shall focus here on
the $q=0$ sector which is relevant in an optical experiment.  As a
consequence of the $C_{2v}$ symmetry, the $q=0$ states can
 be split in four separate symmetry sectors
$(p,i)=(\pm,\pm)$, corresponding to $\pm 1$ characters,
respectively. Although the Raman spectrum will only involve finite
excitations belonging to the $(+,+)$ sector (the Raman operator
bears the full lattice symmetry), it is useful to consider the
four possible symmetry sectors.
Before presenting our results about the four spectra, it is
necessary to briefly discuss the possible connection of the above
quantum numbers $p$ and $i$ defined for the lattice model with
symmetries of the bosonized effective model. First, we note that
{\it parity} corresponds also to a symmetry of the bosonized
hamiltonian and of the Sine-Gordon model. In contrast, under the
above inversion symmetry, the bosonized Hamiltonian is not
invariant by rather transforms according to $D\rightarrow -D$ and
$\theta\rightarrow \theta +\pi$ corresponding to a discrete gauge
symmetry (the notion of ``even'' and ``odd'' sites disappears in
the long wavelength limit). Therefore, the excitations of the
bosonized Hamiltonian can only be classified according to $p=\pm$.
 As we discussed in
Sec.~\ref{sec:strong}, the states in the field theory can only be
characterized by their parity. We recall that we have found that
linear combinations of one soliton and one antisoliton states
could be either odd or even under parity, whereas breathers had to
be even under parity. These simple symmetry-based considerations
will in fact be quite useful for a precise identification of the
various eigenstates found in the numerics.

\begin{figure}[!ht]
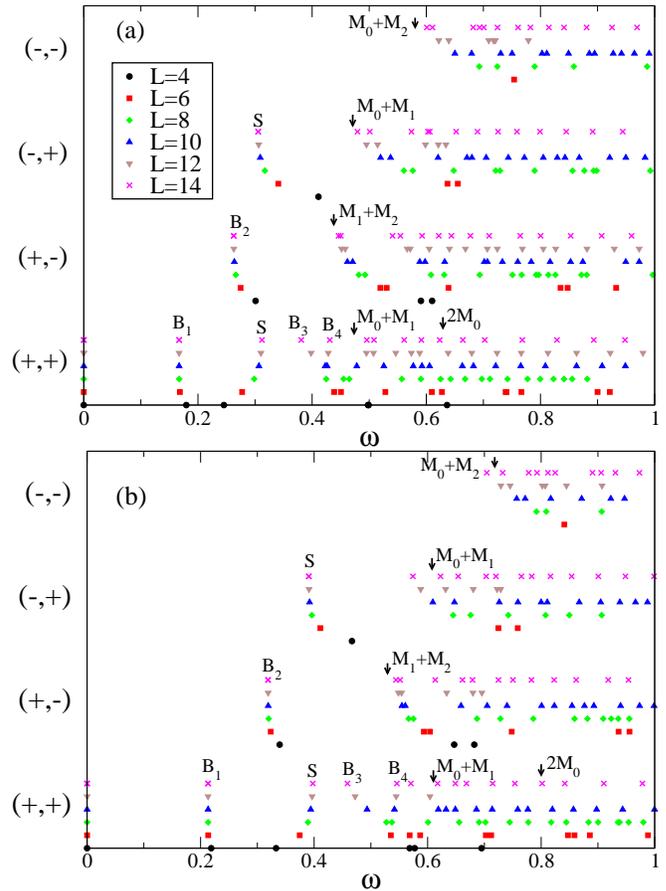

\begin{center}
\includegraphics[width=\linewidth,clip]{fig4a.eps}
\includegraphics[width=\linewidth,clip]{fig4b.eps}
\end{center}
\caption{(Color online) Low-energy spectra in the four
different $(p,i)$ symmetry sectors for $2\times L$ ladders, $L$
ranging from $L=4$ to $L=14$ according to color/symbol codes shown
on the plots, and for $D=0.1$ (a) and $D=0.15$ (b). All
energies are measured from the GS energy ($E=0$). Full (LAPACK)
diagonalization (small size $L=4,6$), Lanczos and/or Davidson ED ($L=8,10,12,14$)
 have been performed~\protect\cite{note_Lanczos}.
The tentative onsets of the 2-particle continua are indicated by arrows.}
\label{fig:spectra}
\end{figure}

\begin{figure}[!ht]
\begin{center}
\includegraphics[width=0.8\linewidth,clip]{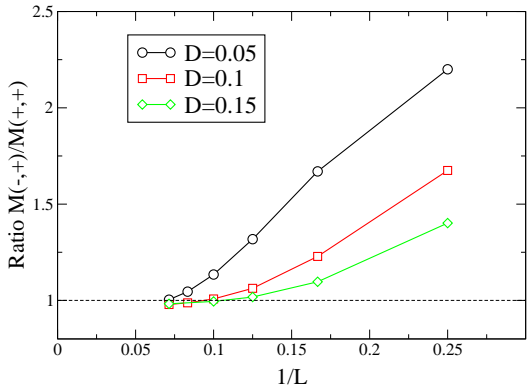}
\end{center}
\caption{(Color online) Ratio of the two quasi-degenerate excited states in the
$(+,+)$ and $(-,+)$ sectors (labelled as ``S'' in
Fig.~\protect\ref{fig:spectra}) as a function of inverse system size for
different values of $D$ (as indicated on the plot). }
\label{fig:ratios}
\end{figure}

We have computed the low-energy spectrum in each symmetry sector
of $2\times L$ periodic ladders with $L=4,6,8,10,12$ and $14$.
Results for $D=0.1$ and $D=0.15$ are shown on
Fig.~\ref{fig:spectra}. At low energy, well-defined discrete
energy levels clearly saturate to constant values for increasing
system sizes. Remarkably, we observe that these states remain
separated from the rest of the spectrum which can be characterized
as a "continuum" from the accumulation of levels appearing with 
increasing cluster size. The discrete levels then correspond
either to the soliton or to the breathers of the previous Sections
to which we now try to make a precise assignment. To do so, we
first notice that two of these levels in the $(+,+)$ and $(-,+)$
sectors are almost degenerate and might correspond to the even and
odd parity soliton excitations (see above). As shown in
Fig.~\ref{fig:ratios}(a), a finite size scaling analysis indeed
proves that the ratio of these frequencies converge to 1 in the
thermodynamic limit so that these levels can indeed be assigned to
the solitons in Fig.~\ref{fig:spectra}.

\begin{figure}[!ht]
\begin{center}
\includegraphics[width=\linewidth,clip]{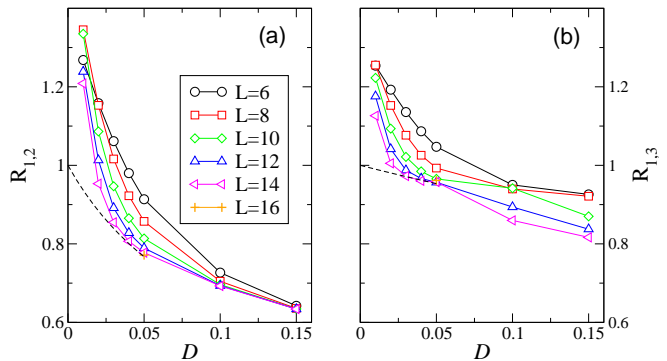}
\end{center}
\caption{(Color online)  $R_{1,2}$ (a) and $R_{1,3}$ (b)
of Eq.~(\protect\ref{eq:relations}) as a function of $D$ for
different systems sizes (as indicated on the plot). The expected
$L\rightarrow\infty$
extrapolation is tentatively sketched as a dashed line.
The magnetic field is such that $m=0.5\, m_{\rm sat}$ and
$D$ is measured in units of $J_\perp$.} \label{fig:relations}
\end{figure}

After having identified the soliton level (of energy $M_0$), we
can also safely identify the lowest energy excitation (which is
well separated from the rest of the spectrum) as the $B_1$
breather of the SG model. We now tentatively assign the remaining
discrete levels as $B_2$, $B_3$ and $B_4$, as shown in
Fig.~\ref{fig:spectra}, and provide a refined analysis that
validates this initial guess. First, we note that all suspected
breathers indeed correspond to $p=+$ as expected from the SG
theory. Secondly, we use the soliton and $B_1$ levels as
``references'' for the identification of the rest of the
low-energy spectrum by performing the following analysis: from
Eq.~(\ref{eq:breather-masses}) of the previous bosonization
approach, one expects that the following combinations of the mass
ratios,
\begin{eqnarray}
R_{1,2}&=&\left(\frac{M_1}{2M_0}\right)^2+\left(\frac{M_2}{2M_1}\right)^2\, ,  \nonumber\\
R_{1,3}&=&\frac{1}{3}\frac{M_3}{M_1}+\frac{4}{3}\left(\frac{M_1}{2M_0}\right)^2\,
, \label{eq:relations}
\end{eqnarray}
converge to $1$ in the limit $D\rightarrow 0$. Such relations can
then be considered as useful ``consistency checks'' of our initial
assignments. We have plotted $R_{1,2}$ and $R_{1,3}$ in
Fig.~\ref{fig:relations} as a function of $D$ and for different
systems sizes. These plots are indeed consistent with the fact
that $R_{1,2}\rightarrow 1$ and $R_{1,3}\rightarrow 1$ for
$L\rightarrow\infty$ and (then) $D\rightarrow 0$, validating the
assignments of the $B_2$ and $B_3$ breathers. As far as $B_4$ is
concerned, its identification is more risky but still very
plausible, at least for $D\ge 0.1$. Lastly, we note that, as seen
in  Fig.~\ref{fig:spectra}, simple kinematic rules can be used to
combine two of the above excitations (summing up their energies)
and lead to the right order of magnitude for the two-particle
thresholds (shown by arrows in Fig.~\ref{fig:spectra}) as well as
the correct quantum numbers..

As a last check, we have considered another (naive) alternative
for the $B_3$ level as the onset of a 2-particle continuum made of
independent $B_1$ excitations. However, the extrapolation of the
$M_3/2M_1$ ratio (not shown) in the thermodynamic limit is
inconsistent with $M_3/2M_1$ being exactly 1 for all values of
$D$. This gives additional credit to our previous assignments and
suggests the absence of a continuum at $2M_1$.

\subsection{Raman scattering and spectral weight}

We now turn to the calculation of the Raman spectrum and investigate the effect of the DM interaction.

It is physically instructive to write the ($T=0$) Raman spectral function in the Lehmann's representation,
\begin{eqnarray}
I_R(\omega)&=&-\frac{1}{\pi}\Im m\langle 0 |\hat{O}_R \, \frac{1}{\omega-H+E_0 + i 0^+} \, \hat{O}_R | 0\rangle\label{eq:propagator}\\
&=&\sum_p | \langle p | \hat{O}_R | 0\rangle |^2 \delta(\omega-(E_p-E_0))  \, ,
\end{eqnarray}
in terms of the physical excited states  $| p\rangle$ of excitation energies $E_p-E_0$.
When the Hamiltonian is isotropic in spin space, since the Raman operator is SU(2) spin-symmetric,
it physically describes singlet excitations or, more picturially,
``double magnon'' excitations. In our case, because of the Hamiltonian anisotropy,
the total spin is no longer a good quantum number and the only remaining
symmetries (apart from the obvious conservation of the full polarisation $m$) are the space group symmetries discussed above.
Operators involving light scattering (like the Raman operator) are translational invariant (on the scale of the lattice spacing)
and so involve only $q=0$ excitations. In addition, for light in-coming and out-going polarizations along high-symmetry directions
(see later), the Raman operator bears the full lattice symmetry
so that the Lehmann sum only contains 
$(+,+)$ excited states. However, the actual spectral weight
might depend strongly on the polarisations. Note that the latter
selection rule exclude the $B_2$ breather from the Raman spectrum.

\begin{figure}[!ht]
\begin{center}
\includegraphics[width=\linewidth,clip]{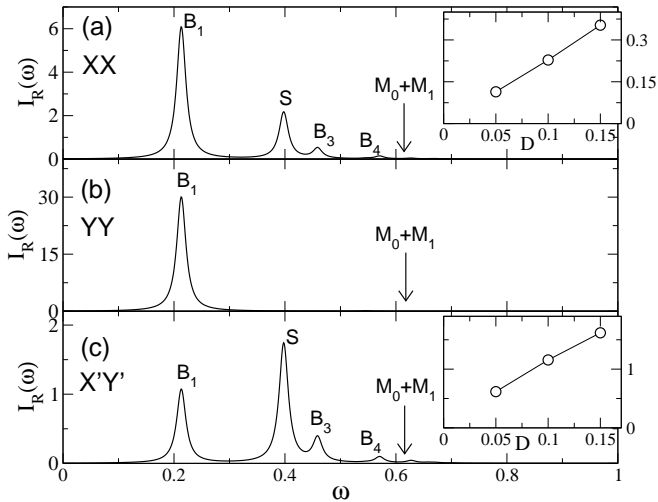}
\end{center}
\caption{Raman spectra, computed on a $L=14$ cluster, for the three different
polarizations XX, YY and X'Y' (see text) and for $D=0.15$ and a magnetic field
such that $m=0.5m_{\rm sat}$.
The spurious $\omega=0$ peak has been subtracted (see text). Note that a small
imaginary part $\epsilon=0.01$
in the frequency has been used to give a small width to the low-energy delta-peaks. The corresponding
weights of these peaks are shown in Fig.~\protect\ref{fig:weights} as
a function of system size. Insets: relative weight of the S state with respect to the B$_1$ one vs $D$, compatible
with a vanishing S intensity as $D\rightarrow 0$}.
\label{fig:Raman_spectra}
\end{figure}

We have used a standard Lanczos technique
to compute $I_R(\omega)$ on finite size clusters up to $L=14$.
Within the Lanczos scheme, dynamical correlations can be conveniently
expressed as simple continued fractions starting from Eq.~(\ref{eq:propagator}).
The spectral weights can also be exactly computed from the coefficients
of the continued fractions. We have checked on small sizes that Lanczos and full
diagonalizations do give identical results.
We have also checked
that, in the absence of magnetic field and DM ($h=0$ and $D=0$),
our results are compatible with Ref.~\onlinecite{natsume_ladder_raman} and
can be simply interpreted as a large weight located at twice the
spin gap (the Raman operator is a spin singlet and light will
create 2 triplet excitations as stated above).

\begin{figure}[!ht]
\begin{center}
\includegraphics[width=\linewidth,clip]{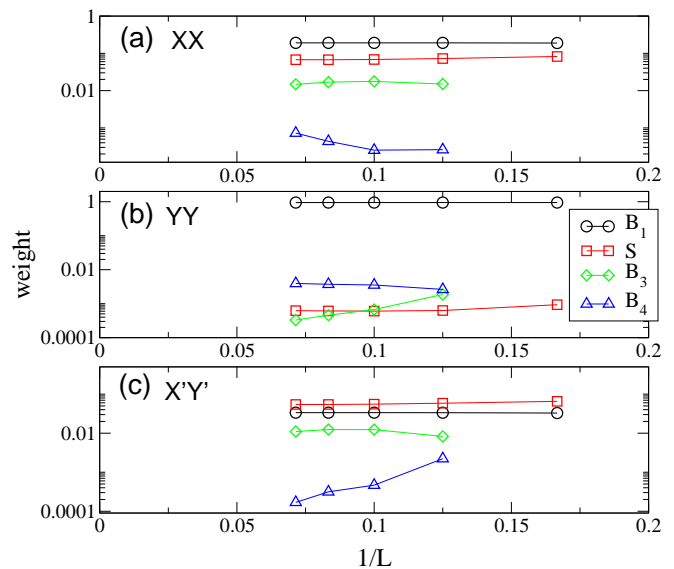}
\end{center}
\caption{(Color online) Finite-size scaling of the Raman spectral weights
of the soliton and the
breathers at $D=0.15$
and for a magnetic field such that $m=0.5\, m_{\rm sat}$.
All light polarisations XX, YY and X'Y' are shown. A logarithmic scale is
used to reveal the tiny weights in the YY polarisation. For XX and X'Y'
polarisations a sizeable fraction of the overall weight (normalized to 1) is
located at higher energies (not shown).}
\label{fig:weights}
\end{figure}

Typical Raman low-energy  spectra are shown in Fig.~\ref{fig:Raman_spectra} for the same parameters as before.
Note that the finite GS expectation values
of the Raman operators produce a peak at zero-frequency. To clarify the plots, we have
subtracted this trivial
contribution (which experimentally is meaningless) and normalized the rest of the spectra so that 
the total integrated weight is unity.
Interestingly, signatures of the soliton and of the $B_1$, $B_3$
and even $B_4$ breathers (all in the $(+,+)$ sector,
see Fig.~\ref{fig:spectra}) can clearly be seen at low frequency in the
XX and X'Y' spectra. However, for YY polarisation, only the $B_1$ peak
is sizeable.
Note also that the low-energy {\it integrated} weight (let us say,
up to $\omega\sim J_\perp\equiv 1$)
depends significantly on the light polarisation. This is directly linked to a 
large transfer of weight to/from the high-energy region, namely
above $\sim 2J_\perp$ (not shown), when changing the light polarisation.
We have also carried out a finite size scaling analysis
of the weights associated to each of the discrete levels, as
shown in Fig.~\ref{fig:weights}. We  also find that (for the same
parameters as before) the relative weight of the S state with
respect to e.g. the B1 breather is 0.35, 0.23 and 0.11 for $D=0.15$,
$D=0.1$ and $D=0.05$ respectively, e.g. in the XX polarization 
(see inset of Fig.~\ref{fig:Raman_spectra}).  This shows that 
the relative weights of the peaks are rather sensitive to the
parameter $D$.  
Our results show small finite size effects and hence strongly
suggest that all the sizeable peaks in Fig.~\ref{fig:Raman_spectra}
survive in the thermodynamic limit.

\begin{figure}[!ht]
\begin{center}
\includegraphics[width=0.9\linewidth,clip]{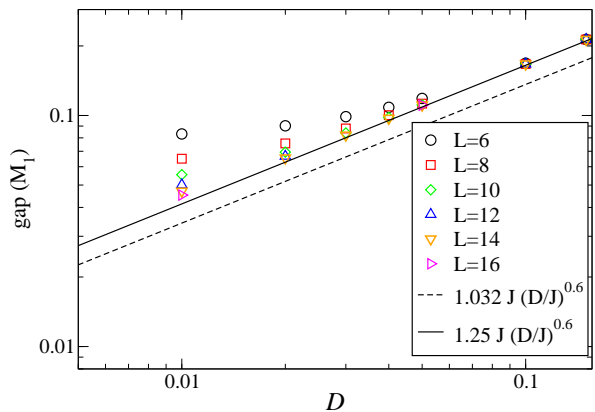}
\end{center}
\caption{(Color online) Evolution of the gap ($M_1$) as a function of
$D$ for a magnetic field such that $m=0.5\, m_{\rm sat}$ (log scale).
The results for several lengths are reported and the
extrapolation for infinite size is consistent with a $D^{3/5}$
behavior (continuous line). The bosonization results shown as
a dashed line differs only by a multiplicative factor of order 1.25.
All energies are in units of $J_\perp$.
}
\label{fig:gap}
\end{figure}

Lastly, we come back to the issue of the gap which is
plotted in Fig.~\ref{fig:magnet}(b). From the previous analysis this gap
corresponds in fact to the excitation energy $M_1$ of the $B_1$ breather.
Its behavior as a function of the DM interaction is shown in Fig.~\ref{fig:gap}
and compared to the bosonization prediction
$\simeq 1.03 J_\parallel (D/J_\parallel)^{3/5}$. Apart from a slightly
different multiplicative factor, the agreement between bosonization and
(extrapolated) numerical results is excellent.

\subsection{Discussion}

To conclude this Section, we would like to discuss in more details the
comparison of the Raman spectra obtained from exact diagonalization
and from bosonization and mapping to the quantum sine-Gordon model.

In the quantum sine-Gordon model, neither operator
$\partial_x \phi$ nor $\cos \theta$ can change the number of
solitons. Thus, they can have non-vanishing 
matrix elements only between the ground
state and the states having an equal number of
solitons and antisolitons and any number of breather. 
As a result, the Raman intensities
predicted from the sine-Gordon model in Sec.~\ref{sec:raman} 
 are in partial disagreement
with the ones obtained from ED if  the state S on
Fig.~\ref{fig:spectra} is identified with a sine-Gordon
soliton. There are three different possibilities to resolve this
discrepancy.  The first one  is that the
state S is actually the second breather B$_2$ of the quantum sine
Gordon model. This is compatible with its symmetry since there is
one such state in the $(+,+)$ sector. This is also compatible with the
observed decrease of the weight of the S peak as $D$ is decreased, as
such a behavior is predicted by  Eq.~(\ref{eq:raman-xx-oe}). 
In that scenario, the ratio of the masses do not agree with
the quantum sine Gordon prediction. The latter result, however,
may result from having relatively large gaps that make 
the continuum limit unjustified. 
However, in this scenario,  we have to understand
why there is a state degenerate with the second breather 
in the sector $(-,+)$.

The second possibility is that the
expression~(\ref{eq:raman-strgcpl}), which is valid for $J_\perp/J_\parallel
\to \infty$ has  corrections of order
$(D/J_\perp)^m$ with $m$ an integer
coming from higher order virtual processes.  
If these higher order
correction terms contain an operator such as $(-1)^n \tau_n^z$ or
$\tau_n^{x,y}$ the bosonized form of which are repectively 
$\cos 2\phi$,  
$\cos \theta \cos 2\phi$ and  $\cos \theta \sin 2\phi$, the Raman
operator will contain soliton creation/annihilation operators.
This scenario could explain the presence of the peak, while
preserving the identification of the state S on
Fig.~\ref{fig:spectra} as a soliton.  Moreover, this scenario is also
compatible with the decrease of the weight in the peak S as $D$ is
decreased. Note however that, in this
appealing scenario, a question remains: why is the B2 breather odd
w.r.t. the {\it inversion} symmetry in contrast to the other
breathers ? Solving this puzzle unfortunately requires
to go beyond the sine-Gordon model (\ref{eq:boso-strong}) 
which does not exhibit such a symmetry.

The third possibility is that the antisymmetric modes discussed in
Sec.~\ref{sec:weak} are also contributing to the spectrum of
Fig.~\ref{fig:spectra} when $D\sim J_\parallel$ 
 even at energy scales small compared to
$J_\perp$. In that case, the continuum field theory
method of Sec.~\ref{sec:strong}
would be useless to understand the spectra as it implicitely
assumes that antisymmetric modes are excited only for energies
comparable with $J_\perp$. 
The fuller theory of Sec.~\ref{sec:weak} would also be of
limited usefulness as the resulting generalized sine-Gordon model
 discussed in \ref{sec:weak}
is non-integrable and thus does not permit to locate breather type
excitations. 

\section{Conclusion}

Although the theoretical interpretation of the Raman intensities
obtained numerically at intermediate coupling in terms of the solitons
and breather of the weak coupling sine-Gordon theory is still a
partially open problem, from the point of view of experiment, 
we can safely conclude that
in a two-leg ladder with Dzialoshinskii-Moriya interaction, the
presence of a gap in the magnetized phase gives rise to bound
states observable in magnetic Raman scattering. In the case of CuHpCl,
earlier studies showed that typical parameters
like $J_\parallel=0.2$ (as used in the numerics) and $D\simeq 0.05$
(all in units of $J_\perp$) give a fair description of the 
magnetization curve~\cite{miyahara06_dm,capponi06_ladder_dm} and 
of the specific heat in an applied magnetic field~\cite{capponi06_ladder_dm}.
Hence, the above investigation of Raman magnetic scattering
(based on analytic and numerical techniques)
clearly establishes that, for light
polarization along the legs, at least two bound states should be visible,
while at least a single bound state should be seen for light polarized
along the rungs. We also expect that the spectral weight of the continuum
predominantly appears  at energies larger than typically 
$\sim 2J_\perp$ i.e., in the case of CuHpCl, above $\sim 3$~meV
(although the ``theoretical'' onset of the continuum is located just above 
the bound states as predicted by the SG description). 
In other words, a ``pseudo-gap'' should form between the bound
states and $\sim 2J_\perp$.
In the particular case of light polarized along the rungs, we further
note that the bound state(s) alone should exhaust most of the overall 
spectral weight. 
We therefore believe that Raman scattering experiments should provide
clear fingerprints of the existence of finite 
Dzialoshinskii-Moriya interaction in low-dimensional magnets.

\acknowledgments
SC and DP are partially supported by the ANR-05-BLAN-0043-01
grant funded by the ``Agence Nationale de la Recherche'' (France). 
SC and DP thank IDRIS (Orsay, France) and  
CALMIP (Toulouse, France) for use of supercomputer facilities.
EO acknowledges hospitality from the university of 
Salerno and support from CNISM  during his stay at the University
of Salerno where part of this research was undertaken.


\begin{thebibliography}{10}
\bibitem{note_soliton}
The ``soliton'' denomination refers to the mathematical property of the
  Sine-Gordon model although the corresponding physical excitation of the
  lattice model does not have here any ``topological'' character.
\bibitem{note_XY} Strictly speaking, this property is only true within the SG
description.
\bibitem{note_Lanczos}
Contrary to the Davidson algorithm, in the Lanczos procedure a careful
  inspection of the evolution of the spectrum with the iteration number is
  necessary to sort out the ``ghosts'' states.
\end{thebibliography}

\end{document}